\documentstyle[12pt]{article}
\advance\textheight by \topskip
\begin{document}
\begin{flushright}
SU-ITP-95-4\\
hep-th/9503029\\
March 5, 1995\\
\end{flushright}
\vspace{-0.2cm}
\begin{center}
{\large\bf     BLACK HOLE  MULTIPLETS AND SPONTANEOUS\\
\vskip 0.6 cm
 BREAKING  OF LOCAL SUPERSYMMETRY}\\
\vskip 1cm
{\bf Renata Kallosh} \footnote {  E-mail:
kallosh@physics.stanford.edu}
 \vskip 0.05cm
Physics Department, Stanford University, Stanford   CA 94305\\
\vskip 0.7 cm

\end{center}
\vskip 1.5 cm
\centerline{\bf ABSTRACT}

\begin{quotation}

We  classify states saturating   a  double  or a single supersymmetric
positivity bound of a four-dimensional $N=4$ supersymmetry.
The  massive four-dimensional double-bound states (Bogomolny states) are shown
to
form a light-like representation of  ten-dimensional supersymmetry. The
single-bound  states form a massive
representation
(centrino multiplet) of a four-dimensional supersymmetry.
The first component of the centrino multiplet  is
 identified  with extreme black holes with regular horizon
which have  one quarter of unbroken supersymmetry. The centrino
multiplet includes a massive spin $3/2$ state, the
centrino,  as a highest spin state.

Existence of massive black hole supermultiplets may affect  the  massless
sector of the  theory. Assuming that gluino condensate is formed one can study
its properties.
The  bilinear combination of covariantly  constant Killing spinors
supplies the possible form for a gluino condensate. The
condensate has null properties, does not introduce a cosmological constant,
and may lead to a spontaneous breaking of local supersymmetry.
This suggests that centrino may provide  a consistent super-Higgs mechanism.

\end{quotation}

\newpage
\section{Introduction}
In this paper we are going to present several different results concerning
physics of extreme black holes in supersymmetric theories. We will begin with
the problem of classification of extreme black holes. This classification
depends crucially on the properties of black holes with respect to
supersymmetry transformations. According to \cite{US}, some of the black holes
are supersymmetric under one quarter of supersymmetries of the original theory.
We will show that they and their superpartners form a massive supermultiplet of
highest spin 3/2. Some other black holes are symmetric with respect to  one
half of original supersymmetries. They form a massive supermultiplet of spin
$1$. Such black holes form a  massive stringy excitations of the
four-dimensional theory. We will show that in fact  they form a light-like
representation from the point of view of the underlying $d=10$ string theory.

Massive supermultiplets of highest spin 3/2 have been studied many years ago,
but then discarded as having no realistic implementations in particle physics,
see, e.g., \cite{strath}. As we will see, these multiplets should be reinstated
in their rights if in addition to particles appearing as elementary field
excitations one also considers extreme black holes and their superpartners. At
this level one can obtain a new example of a consistent theory of massive
particles (solitons) with spin 3/2, which we called centrino. It would be
interesting to consider a possibility of  mixing of centrino and gravitino,
which would make gravitino massive.

Extreme black holes have many different ``faces''.  They can be considered as
point-like objects in $d=4$. However, being uplifted to $d = 10$, some of them
 look like gravitational waves, some may be brought to the form of
gravitational self-dual instantons. Then one can again dimensionally reduce
these solutions to $d = 4$. This can be done  in  different ways, using duality
symmetries and also making different choices for a
four-dimensional manifold. Some of these ways produce  black holes, in some
other ways one can obtain specific gravitational instantons. This reminds the
 relation between static monopoles in four-dimensional Minkowski space
$(t, x^1, x^2, x^3)$
and self-dual instantons in the  four-dimensional Euclidean space  $( x^1, x^2,
x^3, x^4)$ in Yang-Mills theory.

When analysing these configuration we have found that all of them possess a
very interesting property, related to the fact that they admit supercovariantly
constant Killing spinors.

One may try to solve equation of motion for the gluino field in the presence of
extreme black holes. It so happens that the solution for the gluino field can
be easily obtained, and that it does not influence in any way the original
black hole solutions. Thus, extreme black hole solutions admit existence of a
gluino field. This gluino field is analogous in a certain sense to the gluino
condensate introduced in \cite{DRSW} in an attempt to obtain a mechanism of
supersymmetry breaking in string theory. However, unlike the gluino condensate
proposed in \cite{DRSW}, our gluino field,
even without the 3-form condensate, does not introduce any cosmological
constant. This invites us to look very attentively at the black hole physics
from the point of view of stringy phenomenology and spontaneous breaking of
local supersymmetry.

\section{ SUSY bounds}

The ultimate importance of supersymmetric positivity bounds has been realized
by Witten and Olive \cite{WO} in supersymmetric  Yang-Mills theories and
by Gibbons and Hull \cite{GH} in gravitational theories.  After the black hole
solutions saturating the Bogomolny bound of $N=2$ supersymmetry have been found
\cite{GH} there was an extensive study of other solutions of gravitational
theories, which have some kind of unbroken
supersymmetries. The purpose of this section is
to classify  supermultiplets satisfying
either a  single or  a double supersymmetric bound of $N=4$ supersymmetry.
This classification will give some support to the
possibility of having  a massless graviton and  a massive gravitino in
a four-dimensional theory, without explicit breaking of supersymmetry and
without introducing the cosmological constant.

In what follows we would like to classify the known black hole type
configurations
according to the supersymmetry bound (bounds) which they saturate.
 Extended supersymmetries have various representations with
specific properties related to the supersymmetry algebra. An   extensive
review of the representations of extended supersymmetry algebra in various
dimensions was given by  Strathdee \cite{strath}. The supersymmetry algebra
defines
some special representations whose existence is connected with the
saturation of supersymmetric positivity bounds. Some recent studies  of the
monopoles, extreme black holes,  gravitational waves and string states
were associated with such bounds.

In global $N$-extended supersymmetric theories ${N\over 2}$ (${N-1\over 2}$ for
odd $N$)
supersymmetric positivity bounds  arise as an intrinsic property of the
supersymmetry algebra with central charges. The bounds were
explicitly derived by Ferrara, Savoy and Zumino \cite{FSZ} starting from the
following algebra:\footnote{ The notation here are as in \cite{FSZ}.}
\begin{equation}
\{Q_\alpha^i , Q_\beta^j \} = (\gamma^\mu C)_{\alpha \beta} P_\mu \delta^{ij}
+C_{\alpha \beta} U^{ij} + (\gamma_5 C)_{\alpha \beta} V^{ij}\ , \qquad i,j=
1,\dots N\ ,  \qquad \alpha = 1,\dots 4.
\end{equation}
In the Majorana representation $Q_\alpha^i$ are hermitian. The operators
$U$ and $V$ are hermitian and belong to the centre of the superalgebra.
 In Weyl basis in two-component notations the anticommutators
become
\begin{eqnarray}
\{Q_\alpha^i , Q_{\dot \beta}^{j*}\} &=& (\sigma^\mu C)_{\alpha \dot \beta}
P_\mu
\delta^{ij}\ ,\label{s}\nonumber \\
\{Q_\alpha^i , Q_\beta^j \}&=& \epsilon_{\alpha \beta} Z^{ij}\ , \nonumber\\
\{Q_{\dot \alpha}^{i*} , Q_{\dot \beta}^{j* } \}&=& \epsilon_{\dot\alpha
\dot\beta} Z^{*ij}\ .
\label{susyalgebra} \end{eqnarray}
The
matrix $Z^{ij} \equiv -V^{ij} + iU^{ij}$ is complex  skew symmetric. In what
follows we will restrict ourself to the $N=4$ case.
In Weyl
basis the $U(4)$ transformation of supersymmetry charges
\begin{equation}
Q_\alpha^i \rightarrow {\cal U}^{ij} Q_\alpha^j= \tilde Q_\alpha^i\ ,\qquad
Q_\alpha^{*i} \rightarrow {\cal U}^{*ij} Q_\alpha^{*j}= \tilde Q_\alpha^{*}i\ ,
\end{equation}
leaves eq. (\ref{s})
unchanged and forces the central charges to transform like the
complex antisymmetric $[4]_2$ representation of $U(4)$,
\begin{equation}
Z^{ij} \rightarrow ({\cal U} Z {\cal U}^T)^{ij}= \tilde Z^{ij} \ .
\end{equation}
Using this $U(4)$ freedom of a choice of a basis for the supersymmetry charges
one can bring the complex antisymmetric matrix to a normal
form\footnote{It is always possible for one irreducible multiplet, it may
not be possible for few multiplets simultaneously.}
\begin{equation}
\tilde Z = i\sigma_2 \times \left( \matrix{
z_1&0\cr
0&z_2\cr
}\right) \ ,
\label{zmatrix}\end{equation}
where $\sigma_2$ is the Pauli matrix and $\hat Z$ is a $2\times 2$  diagonal
matrix with {\it non-negative real eigenvalues}  $z_1, \; z_2$. After some
additional
symplectic rotation  of  spinorial charges  the supersymmetry
algebra in the rest frame acquires the   following form\footnote{There is one
more
set of operators with the same algebra.}:
\begin{eqnarray}
 \{S_{(1)} , S_{(1)}^{*}\}&=& 2|S_{(1)}|^2 = m - z_1
\geq 0 \ , \nonumber\\
 \{S_{(2)} , S_{(2)}^{*}\}&=& 2|S_{(2)}|^2 = m - z_2  \geq
0 \ ,\nonumber\\
  \{T_{(1)} , T_{(1)}^{*}\}&=&2|T_{(1)}|^2 = m + z_1  \geq 0 \ ,
\nonumber\\
\{T_{(2)} , T_{(2)}^{*}\}&=&2|T_{(2)}|^2 = m + z_2  \geq 0 \ .
\label{z}
\end{eqnarray}
The invariances of the algebra (\ref{susyalgebra}) with non-vanishing central
charges are classified by the orbits of the twofold antisymmetric
representation of $[N]_2$ of $U(4)$. The generic orbit has a little group
$(USp(2))^2$. Thus
in general, for arbitrary central charges the supersymmetry algebra
(\ref{susyalgebra}) is invariant under $(USp(2))^2$ symmetry. In special cases
when
$z_1>z_2$ and $z_2=0$ the critical orbit has a little group  $USp(2)
\times U(2)$. The case  $z_1= z_2$ corresponds to the special critical orbit
with $USp(4)$ invariance.

The two positivity bounds on states in $N=4$ supersymmetry
\begin{equation}
m \geq z_1\ , \qquad m \geq z_2 \ ,
\end{equation}
 exist because these
combinations of the mass and central charges of all states can be
expressed through the square of a particular supersymmetry generators acting
on that state.
The structure of multiplets and its dimension depend crucially on whether
one or both supersymmetric positivity bounds are saturated. Saturation of one
bound inactivates one fourth of the  components of the supersymmetry charge,
leaving the remaining three quarters to generate all states of the multiplet,
starting with the zero helicity state. Saturation of the two bounds leaves only
one half of supersymmetry generators to create the states of the multiplet.
Thus there is a {\it strong discontinuity in the dimension of the
supermultiplet
between the state saturating only one bound which may even be very close to
saturating two bounds
and the state which actually saturate both bounds}. This reminds us of the
situation
in the Yang-Mills theory: small mass still means $3$ polarization states,
whereas the massless state has only $2$.

\section{~Centrino multiplet saturating the single bound of  {} \hskip 1cm {}
$N=4$ SUSY}

Consider the states saturating the single bound  $m = z_1$ with $z_1 >
z_2$.
It is obvious from
eq. (\ref{z}) that saturated bound $m = z_1$ with $z_1 >
z_2$ means that the generators
$S_{(1)}$ drop from the spectrum generating Clifford algebra. Instead of 16
supersymmetry charges we are left with 12. For one half of them we get
\begin{eqnarray}
\label{split1}
 \{S_{(2)} , S_{(2)}^{*}\}&=& 2|S_{(2)}|^2 = z_1 - z_2
 \ ,\\
 \nonumber\\
  \{T_{(1)} , T_{(1)}^{*}\}&=&2|T_{(1)}|^2 =2  m   \ ,\\
 \nonumber\\
\{T_{(2)} , T_{(2)}^{*}\}&=&2|T_{(2)}|^2 = z_1+ z_2   \ .
\label{single}
\end{eqnarray}
The second set of generators has the same structure.

After the rescaling under the strict condition of splitting of the central
charges
$
z_1  \neq  z_2
$
we get
\begin{eqnarray}
\label{split2}
Q_{1} &\equiv  & \left({ m\over z_1 - z_2}\right) ^{1\over 2} S_{2} \ ,\\
\nonumber\\
Q_{2}  &\equiv & \sqrt 2  \; T_{1}   \ ,\\
\nonumber\\
Q_{3} &\equiv & \left( { m\over z_1 + z_2}\right) ^{1\over 2} T_{2}
  \ .
\label{rescale}
\end{eqnarray}
The rescaled generators $Q^i$, $i=1,2,3$,  form  the algebra
\begin{equation}
\{ Q_{i}, Q_{j}^*\}  = \delta_{ij}\, m
\end{equation}
(together with the second set) defining the massive representation without
central charges of $N=3$ supersymmetry repeated twice. The spectrum generating
Clifford algebra can be brought to the $SO(12)$ invariant form with the states
classified by $SU(2) \times USp(6)$. The $SU(2)$ spin will run inside the
fundamental multiplet from $J=0$ up to $J= {3\over 2}$.  We will call the
32-dimensional multiplet with such properties a centrino multiplet. The zero
helicity state will be called a holon  to reflect  the fact that $U(1)^2$
black holes, including Reissner-Nordstr\"om dyons with one quarter of unbroken
$N=4$ supersymmetry provide the zero helicity state for this multiplet
\cite{US}. The spin one-half state will be named holino\footnote{ If we would
consider only the Reissner-Nordstr\"om black holes embedded into $N=2$
supersymmetric theory with one central charge and one half of supersymmetries
unbroken, the corresponding multiplet \cite{GH} would have the highest $SU(2)$
spin
${1\over 2}$ and would consists of the holon and holino and could be called a
holino multiplet. The analogous soliton of global $N=2$ supersymmetry which has
a monopole as the zero helicity state has a spin one-half partner, sometimes
called monopolino and the total multiplet may be called a monopolino multiplet
in accordance with the name of the highest spin state.}. The next partner, spin
one state will be called a centron, to reflect the fact that this massive
representation of supersymmetry carries a central charge. Finally, the highest
spin ${3\over 2}$ state,  centrino  will provide the
name of the total multiplet.

It is quite remarkable that  Strathdee \cite{strath}  gave a detailed
description of this multiplet
and of the quantum numbers of supersymmetry charges and central charges
according to the rest symmetry group $SO(3)\times SU(2) \times SU(2) \times
U(1)$. He also explained that this multiplet contains massive spin
${3\over 2}$
states. Therefore he suggested ``to discard this possibility" and presented his
work in the abstract  as a  ``catalogue of all supermultiplets whose states
carry
helicity $\leq 2$ in the massless cases and $\leq 1$ in the massive case".
Simultaneously in the middle of his work he has  a very important observation
that ``It is not altogether clear what should constitute an acceptable massive
representation but we shall restrict our considerations to those which include
spins $\leq 1$. This reflects the current situation in local field theory  --
at least insofar as theories with a finite number of field multiplets are
concerned  \dots ". On the other hand, it was anticipated in the early days of
supersymmetry by Fayet and Sohnius as well as by
Ferrara, Savoy and Zumino \cite{FSZ}  (who described this multiplet in all
details)  that such multiplets might play a crucial role for a super-Higgs
effect.

The new  attitude to this representation which we are strongly promoting here
came out from the more recent
study of  extreme black holes, which gave a non-trivial example of such
multiplet.
One may  even ignore black holes and simply reinstate  the massive spin
${3\over 2}$ multiplet. One may try to look for this multiplet
directly in particle physics by searching for a  composite-field multiplet.
The most interesting massive multiplets with central charges which  are
represented by  the black
hole solutions have the highest spin state
given by
\begin{equation}
J_{max}  = {N-q \over 2}\ ,
\end{equation}
 where $N$ is the number of supersymmetries, $q \leq {N \over 2}$  ($q \leq
{N-1\over 2}$ for odd $N$)
is the number
of bounds, which are saturated. Thus for $N=2$ we have only one bound
available
and therefore $J_{max} = {2-1\over 2} = {1\over 2}$. For $N=4$ we can
saturate
either two bounds and have $J_{max} = {4-2\over 2}= 1 $  or one bound and have
 $J_{max} = {4-1\over 2}= {3\over 2}$. The known $N=2$ example includes the
Reissner-Nordstr\"om black hole embedded into $N=2$ supergravity
 with  $J_{max} = {2-1\over 2} = {1\over 2}$, holino. The second type
with $J_{max} = 1 $ includes all states saturating a double supersymmetric
bound (Bogomolny bound) of $N=4$ supergravity, in particular pure electric and
pure magnetic dilaton black holes and   axion-dilaton
black holes. They  form a  centron multiplet.
All string states in Sen-Schwarz \cite{SS} spectrum also form centron
multiplets.
The relation between string states and extreme black holes was studied in
\cite{duff}.
Finally, $U(1)^2$ black holes give an example of $J_{max} = {3\over 2}$,
centrino multiplet.\footnote{ For $N=3$ there exists a massive multiplet
(without central charges, $q=0$) with $J_{max} =  {3\over 2}$ and a state
saturating one bound with $J_{max} = {3-1\over 2}=  1$. However, the
 states of $N=3$ supersymmetry have not been identified yet with known
solitons.}

It seems plausible  that when $N=4$ supersymmetry is involved, which typically
has only massless multiplets including gravitons, gluons and photons in field
theory,
the source of the mass parameter can come only from some solitons carrying the
ADM mass of a configuration. This is exactly why before extreme black holes
where embedded into $N=4$ supersymmetry with one quarter of unbroken
supersymmetries  there was no clear or constructive  reason to bring massive
spin ${3\over 2}$ multiplets into the search of a mechanism of  spontaneous
breaking of local supersymmetry.

\section{ Centron multiplet saturating the double bound of $N=4$ SUSY}

If the strong
bound $m = z_1=z_2$ is saturated   the charges $S_{(2)}$ as well as
$S_{(1)}$
drop from the spectrum generating Clifford algebra. We are left with $SO(8)$
invariant Clifford algebra  with the states classified by $SU(2) \times
USp(4)$. The one-particle states are those of massive representation of $d=4,
\, N=2$ supersymmetry algebra without central charges repeated twice. The
$SU(2)$ spin will run from $0$ to $1$.

The algebra of $N=4$ extended supersymmetries without central charges has
$U(4)$ symmetry. The existence of 12 central charges $Z^{ij}$  reduces this
symmetry to the subgroups of $U(4)$ which leave the complex skew
symmetric
numerical $4\times 4$ matrix $Z^{ij}$ invariant. First of all we would like
to stress that not all  representations of four-dimensional $N=4$
supersymmetry algebra with  central charges  can be identified with some
representation of ten-dimensional supersymmetry algebra \cite{FSZ}, but
only the representations with exactly equal $z_1=z_2$.

Even more striking is the fact that the four-dimensional massive
representation saturating the double bound of $N=4$ SUSY can be identified
with  the  light-like representation of the ten-dimensional global
supersymmetry  algebra \cite{FSZ}. Light-like representation are the
massless representation of supersymmetry without central charges. In what
follows, we will explain that the Bogomolny bound can be reexpressed as the
statement that  all states  in ten-dimensional supersymmetric theories
 have
a non-negative mass,
\begin{equation}
m_4  \geq z_1= z_2  \hskip 3 cm  \Longleftrightarrow    \hskip 3 cm
 m_{10}
\geq 0 \ .
\end{equation}
Saturation of the double bound in ten-dimensional language is equivalent to
light-likeness of
the representation,
\begin{equation}
m_4  = z_1= z_2  \hskip 3 cm  \Longleftrightarrow    \hskip 3 cm   m_{10} = 0\
{}.
\end{equation}
In view of the fact that the states saturating the Bogomolny bound have been
under investigation recently,
it might be useful to  present the explanation of the relation above.
Indeed, it is known in principle that $N=4, d=4$ supersymmetry comes from $N=1,
d=10$ supersymmetry. However, only the representations of four-dimensional
supersymmetry with the central charges coinciding  do correspond to the global
representation of $d=10$ SUSY. The argument goes as follows \cite{FSZ}. The
$N=1$ ten-dimensional SUSY algebra is
\begin{equation}
\{Q_\alpha , Q_\beta \} = - (\gamma^\mu C)_{\alpha \beta} P_\mu -
 (\gamma^{A}  C)_{\alpha \beta} P_{A} \ ,
    \qquad  \alpha = 1,\dots 16,   \quad  A= 1, \dots 6,
\label{ten}\end{equation}
where the SUSY charge is a Majorana-Weyl ten-dimensional spinor.
The question is: under which conditions
$
(\gamma^{A}  C)_{\alpha \beta} P_{A}
$
 in some basis can be brought to the four-dimensional form
$
(C_4)_{\alpha \beta} U^{ij} + (\gamma_5 (C_4) )_{\alpha \beta} V^{ij}
$
 so that the ten-dimensional $N=1$ algebra \label{ten}
 coincides with the four-dimensional $N=4$ SUSY algebra with central
charges? To answer this question one has to use the fact that
\begin{equation}
(P^A \gamma_A C) (P^B \gamma_B C)^{\dagger} = P^A P_A \;  {\rm I} \ ,
\end{equation}
where $\rm I$ is a unit matrix.
To get the same diagonal structure from $
(C_4)_{\alpha \beta} U^{ij} + (\gamma_5 (C_4) )_{\alpha \beta} V^{ij}
$
 one has to require that
the matrix of the central charges in the form in eq. (\ref{zmatrix}), to which
it can be brought after appropriate change of the basis, has to be proportional
to the unit matrix
\begin{equation}
\left( \matrix{
z_1&0\cr
0&z_2\cr
}\right) = (P_A P^A )^{1\over 2} \left( \matrix{
1&0\cr
0&1 \cr
}\right)\ ,
\label{nosplit}\end{equation}
which requires $z_1=z_2$. When the four-dimensional mass $m_4$  is equal to the
value of the central charges, one has
\begin{equation}
m_4^2 \equiv  (-P_\mu P^\mu) ^{1\over 2} = (P_A P^A )^{1\over 2}\ , \qquad
m_{10}^2 \equiv (-P_\mu P^\mu) ^{1\over 2} - (P_A P^A )^{1\over 2} =0 \ .
\end{equation}
In conclusion, the centron supermultiplet satisfying a Bogomolny bound of
$N=4$
supersymmetry with the highest spin state $1$  in four dimensions, can be
reinterpreted
as the light-like representation of $d=10, N=1$ global SUSY.  The deep relation
between supersymmetric gravitational pp-waves and uplifted into ten dimensions
extreme black holes
\cite{BKO} holes may be related to this property of the supersymmetric
multiplets.

 Thus we have also learned that centrino multiplet   of the four-dimensional
SUSY, which extends the range of a spin up to ${3\over 2}$, does not form any
representation of the ten-dimensional global SUSY. Indeed,  the mere existence
of the centrino is based on the central charges split (\ref{split1}),
(\ref{split2}). Ten-dimensional global SUSY
forbids such split, according to eq. (\ref{nosplit}). However, centrino is a
perfect representation of $d=4$, $N=4$ supersymmetry.

We would like to use the existence of massive soliton supermultiplets in
extended supersymmetries for the purpose of spontaneous breaking of local
supersymmetry.
In what follows we will discuss one possible way of thinking about it.

\section{ Gluino condensation induced by Killing spinors}
 All known  Lagrangian theories of local $d=4$, $N=4$ supersymmetry describe
only  massless supermultiplets. However, supersymmetric gravitational
solitons (solutions of the full non-linear field equations)   do bring into the
theory mass parameters via the ADM mass. Some of these configurations which
admit supercovariantly constant spinors form the massive supermultiplets
discussed above.

The purpose of this section is to show that black holes with unbroken
supersymmetries  may be capable of
affecting the massless sector of
$N=4$  theory by  inducing a gluino  condensate without a
cosmological constant and therefore (at least some of those solutions) may
supply their  Killing spinors for  consistent  spontaneous symmetry breaking.
We will present here the  observation about all configurations with unbroken
$N=4$  supersymmetry which may be useful for the investigation of their
possible role  in particle physics in connection with spontaneous breaking of
local supersymmetry.

 Suppose that   extreme black holes (or their instanton counterparts which
appear after uplifting black holes to d=10  and then reducing them back to
Euclidean d = 4 space) may induce the gluino condensate. One can then address
the following question:
if some dynamical mechanism exists for such condensation, what are the
properties of the condensate
in this theory?

The  fundamental  property of the Killing spinors admitted by extreme black
holes
in asymptotically flat space-times is the fact that they do not fall off at
infinity. Near the horizon they behave differently, depending on the
configuration
and on whether one works in canonical or stringy frame. In particular, some of
 extreme black holes have Killing spinors vanishing at the horizon in canonical
four-dimensional geometry, some of them have Killing spinors
remaining  constant in the stringy frame. Some extreme black holes
 in canonical four-dimensional geometry have the singularity hidden by the
horizon, some have singular horizons, and some exhibit naked singularities.
However, those with singular horizons and/or naked singularities have a
vanishing area of the horizon.
Because of these various unusual properties there is not much information
available at the moment about the  normalizable fermion zero modes  in the
extreme black hole backgrounds. The systematic study of such fermion modes
would be most desirable and would clarify the issues which are raised in this
paper.

 Most of the solutions with
unbroken $N=4$ supersymmetries in four dimensions has been uplifted to ten
dimensions and  most of the $N=1$ ten-dimensional
supersymmetric solutions have been dimensionally reduced down to four. The
way up and down is not completely unique, there are duality transformations
on the way. Taking into account this ambiguity one may still have a clear
identification of solutions in ten and four dimensions depending on whether
they saturate
a single or a double supersymmetric bound, i.e. whether they have one half or
one quarter of unbroken supersymmetries.

We have found that Killing spinors of ten-dimensional supergravity
configurations  automatically solve the gluino field equations in the
background given by the corresponding configuration. To establish that, one
may consider, e. g., the following combination of equations for the Killing
spinors of ten-dimensional supergravity:\footnote{We are using the
description
of ten-dimensional theory in  \cite{BR}. For the time being we consider
only the classical supergravity coupled to Yang-Mills theory.}
\begin{equation}
\gamma^a \delta \psi_a  + 2 \sqrt 2 \delta \lambda = 0 \ , \qquad
a=0,1,\dots ,9.
\end{equation}
 Using the explicit expression for the variation of the ten-dimensional
gravitino
\begin{equation}
\delta \psi_a = D_{a +} \epsilon \equiv [D_a (\omega) -
{3\sqrt 2\over 8} H_{abc} \gamma^{bc}]\, \epsilon (x)
\end{equation}
 and dilatino
\begin{equation}
\delta \lambda = [- {3\sqrt 2\over 8} \phi^{-1}
\gamma^a \partial_a \phi +
{1\over 8} H_{abc} \gamma^{abc}]\, \epsilon (x)
\end{equation}
 (with all fermions vanishing) one
can verify that the combination above is proportional to  the part of
equation of motion for the gluino in the external field of the uplifted black
hole, which is linear in the gluino,
\begin{equation}
[\gamma^a D_a (\omega) - {3\over 2} \phi^{-1} \gamma^a \partial_a \phi
- {\sqrt 2 \over 8} \gamma^{abc} H_{abc}] \, \chi (x) =0\ .
\end{equation}

The part of the gluino equation coming from the the 4-gluino term is
proportional to
\begin{equation}
 \gamma^{abc}  \chi \;  tr(\bar \chi \; \gamma_{abc} \chi) \ .
\label{3gluino}\end{equation}
This term will vanish either when gluino  belong to an
Abelian group or as a consequence of some constraints.  For example, if the
non-Abelian gluino satisfies the light-cone constraint, which we will discuss
later, this term will drop from the equations of motion for gluino. One also
has to
take into account that in the presence of gluino the supersymmetry variation of
the
gravitino,  as well as of the dilatino,  is modified according to \cite{BR},
\begin{eqnarray}
\delta_\beta  \psi_a &=& {1\over 192} \beta \;  \gamma^{bcd} \Gamma_a
\epsilon (x) \; tr (\bar \chi \; \gamma_{bcd} \chi)  \ ,\\
\nonumber\\
\delta_\beta \lambda &=& {1\over 384} \beta \; \gamma^{bcd}  \epsilon(x) \;
tr(\bar \chi \; \gamma_{bcd} \chi) \ ,
\end{eqnarray}
where $\beta$ is a constant in front of the Yang-Mills multiplet.

We would like to find a solution of the gluino field equations using the
Killing spinors
of the configuration without fermions first. Note that the gluino has the same
ten-dimensional chirality as the parameter of the supersymmetry
transformations. Afterwards we may investigate whether we still have Killing
spinors in the presence of the non-vanishing gluino. We may further  look for
the
gluino condensate using the Killing spinors to construct  the bilinear
combination of gluinos. We will study various possibilities for  different type
of configurations admitting Killing spinors.  Finally the effect of the gluino
condensate on the local supersymmetry transformations of the gravitino and
the dilatino can be studied.

The main feature of any uplifted extreme black hole configuration is the number
and the form of the linear constraints which the Killing spinors satisfy.
Besides, the uplifted black holes depend only on $x^1, x^2, x^3$. Therefore
they form a subset of all configurations which have unbroken supersymmetries in
ten dimensions. For simplicity we will call them uplifted black holes with the
understanding that as the ten-dimensional configuration some of them may look
as pp-fronted
gravitational waves or instantons of the four-dimensional submanifold or
monopoles.

The large set of ten-dimensional supersymmetric configurations includes
supersymmetric waves, fundamental strings, dual waves, chiral null models,
etc. They may also depend on $x^4, x^5, x^6, x^7, x^8$, but typically they are
independent of $x^0$ and $x^9$. Some of them are related to uplifted
 electrically charged black holes. The corresponding Majorana-Weyl
ten-dimensional Killing spinor
satisfies the light-cone condition
\begin{equation}
\gamma^+ K \equiv  (\gamma^0 + \gamma^9 )K = 0 \ .
\label{killing1}\end{equation}
Magnetically charged uplifted  black holes usually are constrained
by the chirality
condition
\begin{equation}
(1+ \gamma^5) K \equiv (1 + \gamma^1 \gamma^2 \gamma^3 \gamma^4)K =0 \ .
\label{killing2}\end{equation}
In both cases the number of Killing spinors is equal to $8$ since both the
light-cone constraint (\ref{killing1}) as well as the chiral constraint
(\ref{killing2}) break any spinor into two parts and
the  unconstrained spinor is $16$-dimensional. Pure electric configurations can
be brought to pure magnetic ones by the $SL(2,R)$ rotation. In ten-dimensional
geometry this corresponds to a rotation of the Killing spinor from the
light-cone one to the chiral one. Both are eight-dimensional.

Uplifted $U(1)^2$ black holes  with one quarter of unbroken supersymmetry
satisfy both
the light-cone as well as chirality
constraints.\footnote{Our four-dimensional
choice of electric and magnetic $U(1)$ directions was  $\alpha^3$ for the
vector field and $\beta^3$ for the axial-vector field \cite{US}. This
corresponds
to  $x^6$ for the
electric field and $x^9$ for the magnetic one. We prefer to use as an electric
direction the $x^9$-coordinate  and as the magnetic one the
$x^4$-coordinate. With such a choice the meaning of the one quarter of
unbroken supersymmetry becomes very clear.} The first condition
leaves one half of the 16-component ten-dimensional spinor, the second
condition leaves one half of the remaining 8 components.   Thus
in any case we get a 4-component Killing spinor. In the limit that the
magnetic charge vanishes, we get only the first light-cone constraint and the
Killing spinor becomes  an 8-component light-cone spinor, like the Killing
spinor of the ten-dimensional pp-fronted supersymmetric waves. When the
electric charge vanishes, we get an 8-component Killing spinor which is chiral
in 4-dimensional Euclidean manifold $x^1, x^2, x^3, x^4$. This is a
standard property of   Killing spinors of the monopoles,  instantons, and
uplifted magnetic black holes. However, as long as both charges of two
different $U(1)$ fields are present in the solution, which provides the split
of central charges, the Killing spinor has to satisfy both constraints, given
in
eqs. (\ref{killing1}), (\ref{killing2}). The torsionful curvature of the
ten-dimensional space has both some null properties in $u=t-9$ direction and is
self-dual  in the four-dimensional Euclidean space $1,2,3,4$.

Having established the constraints (\ref{killing1}), (\ref{killing2}),
which are satisfied by
Killing spinors,  we may proceed with the investigation of the 3-gluino
term (\ref{3gluino}). The linear part of gluino field equations is solved by
Killing spinors
for any non-Abelian gluino. Therefore in the first approximation we can take
any of the components the non-Abelian gluino field $\chi$ to be proportional to
a
 Killing spinor  $K$. However, if we would like also to get rid of the 3-gluino
term
(\ref{3gluino}) we may either rely on the light-cone constraint for gluino,
which makes  the 3-gluino term vanishing, or just choose only an Abelian gluino
field. The 3-gluino term vanishes for any Abelian ten-dimensional Majorana-Weyl
spinor. This second possibility also makes the corrections to the Killing
spinor equations due to gluino
vanishing.  Having all these options in mind we may  look for the gluino
condensate using the Killing spinors to construct  the bilinear combination
of gluinos.

All black holes  which admit Killing spinors have horizons, when considered
in canonical $d=4$  geometry. In general, we are not aware of supersymmetric
solitons in the theory of gravity in  asymptotically flat four-dimensional
space
which are free of horizons and  singularities. Rather, one may divide all known
solutions as to whether the horizon is singular or whether it is not and
covers
the singularity  hidden behind the horizon. The first type of the solutions
seems mostly   related to the states  saturating the strong Bogomolny
bound (double bound of $N=4$ supersymmetry).\footnote{
It would be very interesting to find solutions, saturating the double bound
with
non-singular horizon in canonical four-dimensional  geometry.} In particular,
pure electric or pure magnetic $a=1$ or $a=\sqrt 3$  dilaton black holes have
singular horizon with vanishing area in canonical geometry. All
such solutions
have one half of unbroken $N=4$ supersymmetry, i.e. the Killing spinor
has  8  components.

On the other hand, known $U(1)^2 $ black holes
 \cite{US} have 4
Killing spinors which
 tend to a  constant far away from the
black hole and vanish at the horizon. In addition, these black holes have
regular horizon in canonical four-dimensional geometry. The singularity of
these extreme black holes is behind the horizon as long as the central
charges are split, $z_1 > z_2$, since the singularity is  at $r= z_2 $ whereas
the event horizon is at
$r_+ = m =z_1 >z_2$. These solutions saturate only a single
bound of $N=4$ supersymmetry.

  As a simplest possibility we may take an uplifted
Reissner-Nordstr\"om dyon with equal electric and magnetic charges,
$m=z_1, \; z_2 =0$.   This particular solution has no dilaton,
the stringy frame coincides with the canonical one  in  four as well as
in ten dimensions. From the uplifted anticommuting covariantly constant Killing
spinors
$K =\sqrt { 1- {m\over r}} \;  \epsilon_0$
 satisfying the constraints  (\ref{killing1}), (\ref{killing2})  one can
construct their quadratic combination which may serve as a possible form of
 gluino condensate  as follows:
\begin{equation}
\aleph^{abc} \equiv tr (\bar \chi \; \gamma^{abc} \chi )=
C\;  \bar K \; \gamma^{abc} K= C\;
(1- {m\over r}) \bar \epsilon_0 \;
\gamma^{abc} \epsilon_0\ , \qquad a = 0,1,\dots 9.
\end{equation}
Here $\gamma^{abc}$ is a completely antisymmetrized product of
three gamma matrices in the tangent space, $\epsilon_0$ is a constant spinor,
satisfying both constraints
(\ref{killing1}), (\ref{killing2}). The dependence of the condensate
on the distance from the black hole is $(1- {m\over r})$,\, $r^2 = (\vec x
)^2 = (x^1)^2 +
(x^2)^2 + (x^3)^2$; it comes from the dependence of the Killing spinors on
the geometry, . The constant $C$ is introduced to take care
of the dimension of the gluino versus the Killing spinor.

In isotropic coordinates which at positive $\rho= r - m $ describe only the
space outside the horizon we have
 \begin{equation}
\aleph^{abc}  = C\;
\left (1+ {m\over \rho}\right )^{-1} \bar \epsilon_0 \;
\gamma^{abc} \epsilon_0\ , \qquad a = 0,1,\dots 9.
\end{equation}
When $\rho \rightarrow \infty$ the condensate tends to a constant, near the
horizon when $ \rho \rightarrow  0$ the condensate vanishes. For the multi
black hole solutions we get
\begin{equation}
\aleph^{abc}  = C\;
\left (1+ \Sigma_s  {m_s \over  |\vec x - \vec  x_s|}\right )^{-1} \bar
\epsilon_0 \;
\gamma^{abc} \epsilon_0\ , \qquad a = 0,1,\dots 9,
\end{equation}
where $\vec x_s, s= 1, \dots , n$ is the position of  each black hole.

The non-vanishing components of $\aleph^{abc}$ can be
chosen to be
\begin{equation}
\aleph^{- st} \equiv tr(\bar \chi \; \gamma^{-st} \chi) ,    \qquad
\gamma^{-} = \gamma^0 - \gamma^{9}, \quad  s,t=5,6,7,8.
\end{equation}
The reason for this choice is the following. To satisfy the light-cone
constraint
(\ref{killing1}) we have to choose at least one out of $\gamma^a, \gamma^b,
\gamma^c$ to
be
$\gamma^-  $ and the rest could be in any part of $SO(8), a= 1,\dots , 8$.
 For the rest we take into
account that  we have only a four-component anticommuting Killing spinor,
therefore there exists a six-component bilinear combination of such spinors.
Since we are interested in a combination whose vacuum expectation value  does
not vanish, we choose out of all possible $SO(8)$ antisymmetric tensors only
the part of it belonging to $SO(4)$ since directions $5,6,7,8$ are flat. The
equivalent statement about the condensate is the following. There exists a
covariantly constant bilinear combination of the covariantly
constant  Killing spinors
\begin{equation}
\aleph _{[\alpha \beta] } \equiv  C\; K_\alpha K_\beta \ , \qquad D_{\mu +}\;
(\aleph _{[\alpha \beta] })=0 \ ,
\end{equation}
which provides the gluino condensate.
This is a natural choice for the Reissner-Nordstr\"om dyon with one quarter of
unbroken supersymmetries.

This choice is acceptable also for all solutions with light-cone Killing
spinor  with
 one half of unbroken supersymmetries.  The square of the
gluino condensate,  which is build out of light-cone anticommuting Killing
spinors, vanishes since the condensate has a null property. Namely, since the
component
$\aleph^{+ st}$ vanishes due to the light-cone constraint (\ref{killing1}) we
have
\begin{equation}
(\aleph^{abc})^2  \equiv \left (tr (\bar \chi \; \gamma^{abc} \chi )\right)^2 =
 tr (\bar \chi \;
\gamma^{-st} \chi) \;  tr (\bar \chi \;
\gamma_{-st} \chi) = 0 \ .
\label{lambda}\end{equation}
The contribution to the gluino field equation from the 4-gluino term also
vanishes, since
\begin{equation}
tr (\bar \chi \;\gamma^{-st} \chi)  \;
\gamma_{-st} \chi=0 \ .
\end{equation}
For magnetic solutions one can also use this choice, by choosing only those
of 8 Killing spinors which satisfy the light-cone constraint.

 Thus the  gluino condensate may appear in the supersymmetry
transformation of the gravitino and dilatino and may lead to spontaneous
violation of supersymmetry. Also it may induce a mass for the  gravitino
via the
2-gravitino-2-gluino-interaction terms. There will be no cosmological
constant\footnote{The mechanism of gluino condensation
studied in \cite{DRSW} was
developed for the Calabi-Yau manifold. It was necessary there to compensate
the gluino condensate by a 3-form condensate, since there was only one
non-null possible structure  related to an $SU(3)$ holonomy of the six
manifold.}
due to
the condition (\ref{lambda}). A possible scenario is the following.  One may
start with the
action  of the ten-dimensional supergravity  coupled to Yang-Mills fields.
Since
the Killing spinor, which does  not vanish at infinity, solves the equations of
motion of gluino, we may shift the  gluino field  on the Killing spinor.
\begin{equation}
\chi \rightarrow  \chi + C\: K \ .
\end{equation}
The new gluino will be a well behaved   field and appropriate for
  the standard formulation of the quantum field theory (i.e. it will have the
behavior at infinity
which is appropriate for a field over which one integrates in the path
integral). This  reminds us of  the situation with the
separating the non-vanishing at infinity  vacuum expectation of the Higgs
field. After this procedure we will get in the local supersymmetry
transformation rules for
gravitino and dilatino all kind of usual terms, depending on fields of the
theory which fall off at infinity and vanish near the horizon. The exceptional
terms will be
\begin{eqnarray}
\delta  \psi_a &=& \dots  + {1\over 192} \beta \;   \gamma^{bcd} \Gamma_a
 \epsilon (x)   C\; \left (1+ {m\over \rho} \right )^{-1} \; tr (\bar
\epsilon_0 \; \gamma_{bcd}
\epsilon_0) \ , \\
\nonumber\\
\delta \lambda &=& \dots +{1\over 384} \beta \; \gamma^{bcd}
\epsilon(x) \; C\;  \left (1+ {m\over \rho} \right )^{-1} \; tr (\bar
\epsilon_0 \; \gamma_{bcd}
\epsilon_0)\ .
\end{eqnarray}
Here $\epsilon(x)$ are generic local supersymmetry parameters depending
 on ten-dimensional coordinates, $\epsilon_0$ are the constant spinors
representing the Killing spinors far away from the black hole.

It would be  interesting to relate this observation to the properties of the
moduli
space of the system of two black holes. It was found in \cite{GK} that the
Euler
number for the  Reissner-Nordstr\"om case  equals $1$,
which may be a signal of spontaneous breaking of supersymmetry. Indeed, if one
interprets the Euler number of the moduli space manifold as the Witten index,
one discovers that the supersymmetry is broken.
For  a  two-black hole system with $a^2=1$
 the moduli space has a conical singularity, and for $a^2=3$ the
 moduli space is  flat.

In conclusion, we suggest to use the
 centrino  as well as  the centron multiplets to break local supersymmetry, for
example by
inducing   gluino condensation. From the point of view of the ten-dimensional
stringy geometry
magnetic type centrons with the first component of the multiplet given by the
magnetically charged black holes are most attractive:  stringy
$\alpha'$-corrections can be taken care of by embedding the spin connection
into
the gauge group.
Electric centrons also have advantages since at least some of them belong to a
family of supersymmetrized pp-fronted
ten-dimensional gravitational waves and also permit  the embedding of the spin
connection into the gauge group. Note that  magnetic and electric charges of
these configurations are  the central charges of the supersymmetry algebra.

{}From the point of view of a four-dimensional canonical geometry, centrino
multiplets  related to Reissner-Nordtsr\"om-type  $U(1)^2$ dyon black holes
\cite{US}  seem to deserve special attention.  Indeed,   they contain  a
massive spin ${3\over 2}$
state (without a massive spin $2$ state). It would be most interesting to build
the effective action for this multiplet and investigate the possibilities to
mix  it with gravitino. The fact that the natural structures for the
gluino condensation are provided
by the centrino black hole looks encouraging. However,  much more work will be
required to understand whether indeed all consistency requirements for the
super-Higgs mechanism can be met. The  scenario of spontaneous breaking
of local supersymmetry via massive supersymmetric configurations with central
charges outlined above needs to be
supported by a specific dynamical mechanism.

\vskip 0.5 cm

I am very grateful  to  R.
Brooks, A. Linde, T. Ort\'{\i}n, S.-J. Rey and L. Susskind for many fruitful
discussions of the issues studied in this paper.

\end{document}